\documentclass[12pt]{iopart}
\pdfoutput=1
\usepackage{hyperref}
\hypersetup{colorlinks, allcolors={blue}}
\usepackage{subcaption, graphicx} 

\renewcommand{\tref} [1]{Tab.~\ref{#1}}

\renewcommand{\fref} [1]{Fig.~\ref{#1}}
\renewcommand{\Fref} [1]{Figure~\ref{#1}}

\newcommand{\cref} [1]{ref.~[\onlinecite{#1}]}
\newcommand{\Cref} [1]{Reference~[\onlinecite{#1}]}

\begin{document}
\title[]{Electronic properties and transport in metal/2D material/metal vertical junctions}
\author{Ga\"{e}lle Bigeard$^1$, Zineb Kerrami$^{1,2}$, Fran\c{c}ois Triozon$^2$ and Alessandro Cresti$^1$}
\address{$^1$ Univ. Grenoble Alpes, Univ. Savoie Mont Blanc, CNRS, Grenoble INP, CROMA, 38000 Grenoble, France\\
$^2$ Univ. Grenoble Alpes, CEA, Leti, F-38000 Grenoble, France.}
\vspace{10pt}
\begin{abstract}
\\
We simulate the electronic and transport properties of metal/{\mbox{two-dimensional}} material/metal vertical heterostructures, with a focus on graphene, hexagonal boron nitride and two phases of molybdenum diselenide. 
Using density functional theory and {\mbox{non-equilibrium} Green’s function}, we assess how stacking configurations and material thickness impact important properties, such as density of states, potential barriers and conductivity. 
For monolayers, strong orbital hybridization with the metallic electrodes significantly alters the electronic characteristics, with the formation of states within the gap of the semiconducting 2D materials. 
Trilayers reveal the critical role of interlayer coupling, where the middle layer retains its intrinsic properties, thus influencing the overall conductivity. 
Our findings highlight the potential for customized multilayer designs to optimize electronic device performance based on \mbox{two-dimensional} materials.
\end{abstract}
\vspace{2pc}

\section{Introduction}
The pursuit of sustainability and the reduction of power consumption are becoming central considerations in the development of electronic devices in a world where data storage and logic devices are experiencing exponential growth~\cite{IIRDS2023}. 
To address this issue, several solutions have been proposed, such as the use of advanced materials, innovative fabrication methods to reduce their size, and the reduction of the consumption of energy for logic devices. 
This can be achieved, for example, through tunnel \mbox{field-effect} transistors~\cite{Avci2015} or \mbox{three-dimensional} architectures, which facilitate the vertical integration of logic and memory on a single chip, thus reducing the energy consumption associated with data transfer. 
In this regard, \mbox{two-dimensional}~(2D) materials are highly suitable for the so-called \emph{beyond CMOS} technology~\cite{Robinson2018,Briggs2019,Dragoman2021,Hao2022,Cresti2023}. 
Thanks to their \mbox{atomic-scale} thickness, 2D materials offer unique electronic and physical properties not found in their bulk counterparts, including reduced material usage, easier electrostatic control with lower energy consumption and tunable electronic properties, thus making them ideal candidates for next-generation electronic applications. 
In addition, their \mbox{dangling-bond-free} surface considerably reduces interfacial defects and allows van der Waals (vdW) stacking, thus enabling different heterostructures with customized properties for both vertical and horizontal devices, and opening the way for novel physical exploration and advanced electronic devices~\cite{Lemme2022}. 

Among these innovations, atomristors (a term that refers to memristors composed of atomically thin materials) have gained significant interest due to their potential for \mbox{ultra-scaled} \mbox{non-volatile} memories~\cite{Ge2017} with potential neuromorphic computing applications. 
Such devices, which are based on \mbox{metal-insulator-metal} heterostructures, can be used in \mbox{radio-frequency} switches for telecommunications~\cite{Kim2022} and \mbox{on-chip} memories. 
They promise \mbox{low-power} switching in both the \mbox{low-resistance} state (LRS) and \mbox{high-resistance} state (HRS)~\cite{Kim2022}. 
However, the full comprehension of the working mechanisms of such devices remains unclear, and their optimization in terms of switching time, capacitance, ON/OFF current ratio, yield and endurance is still incomplete. 
The scientific community appears to have reached a consensus that LRS is activated by the formation of conductive filaments between electrodes, due to defects and/or substitution of 2D material atoms with metallic atoms from contacts~\cite{Hus2020,Shah2023,Shah2024,Mitra2024}. 
The nature of the filaments can vary: migration of metallic atoms~\cite{Hus2020,Shah2023,Shah2024}, 2D material vacancies~\cite{sangwan_gate-tunable_2015, chen_introduction_2023, sangwan_multi-terminal_2018}, or oxygen vacancies~\cite{wang_robust_2018, jeong_graphene_2010, krishnaprasad_mos2_2022}, and so their creation and destruction can be driven by different processes (such as electric field or Joule heating for bipolar or unipolar switching, respectively). 
It is also possible that other phenomena than conductive bridge formation are at work in these devices, such as structural phase transitions~\cite{zhang_electric-field_2019}. 
The HRS depends on the vertical transmission through the \mbox{metal-2D} material interfaces in the absence of conductive filaments. 
Therefore, the nature of the 2D materials and contacts, their thicknesses and stacking orientation are expected to play a central role. 
Understanding these effects could provide valuable insight into how interfacial interactions impact overall performance and efficiency, thus improving the design and optimization of these advanced electronic devices. 

The purpose of this study is to provide insight into the interplay between the intrinsic properties of 2D materials, their thickness and the metal/2D material interface, and on how these factors collectively influence the \mbox{out-of-plane} transport properties. 
We simulate the structural, electronic and transport properties of monolayer and multilayer 2D materials sandwiched between two metallic electrodes. 
We consider four 2D materials, namely \mbox{semi-metallic} graphene (Gr), insulating hexagonal boron nitride (hBN), molybdenum diselenide in its semiconducting phase \mbox{(2H-MoSe$_2$)} and metallic phase \mbox{(1T-MoSe$_2$)}, and nickel (Ni) (111) as the metallic contact. 
Our findings suggest that the first 2D layer at the interface with \mbox{non-passivated} contacts is bonded to the metallic electrodes, thus showing a metallic nature, the characteristics of which strongly depend on the specific stacking.

\section{Methodology}
We simulate a complete structure composed of a 2D material within thick top and bottom electrodes with \mbox{in-plane} periodicity, see \fref{fig1}. 
The relaxed geometrical structure and its electronic structure are calculated with density functional theory (DFT), as implemented in the SIESTA code~\cite{Garcia2020} on a localized basis set. 
The nonequilibrium Green's function technique, as implemented in SIESTA through the TRANSIESTA and TBtrans codes~\cite{Papior2017}, is then employed to calculate the current flowing through such devices. 
We perform data \mbox{post-processing} with the SISL Python library~\cite{Papior2017}. 
As detailed below, the simulation parameters, including grid densities, basis optimization and material constraints, are carefully selected to clearly define the simulation environment and tolerances. 

\subsection{Simulation parameters}
We make use of the PBE functional \cite{Perdew1996} and validated optimized norm-conserving Vanderbilt pseudopotentials, sourced from the PseudoDojo website \cite{Hamann2013,van_setten_pseudodojo_2018}. 
The Fermi distribution is used with a smearing temperature of 0.05~eV. 
The relaxation process is carried out exclusively in the \mbox{out-of-plane} direction with a tolerance of 0.002~eV/\AA$^3$ for stress and 0.01~eV/\AA~for atomic forces. 
The vdW corrections due to London dispersion forces are not included in this study, since the coupling between metal and 2D material in our heterostructures is dominated by strong coupling at the interfaces, rather than weak vdW forces. Similarly, the \mbox{spin-orbit} coupling (SOC), while inducing a \mbox{spin-orbit} splitting of especially the valence bands for TMDs with heavy metals, is not expected to significantly affect vertical transport and does not change our conclusions. 
Another motivation to exclude SOC correction lies in their computational expense, as they significantly increase the required resources without substantially altering the relevant outcomes of this study. 
The simulation parameters are chosen to have a final tolerance of 10~m\AA~on distances, $5\times 10^{-3}$~e/\AA$^3$ on the electronic density, 100~meV on the electrostatic potential.
The selected parameters are reported in the Appendix. 
We simulate transport near \mbox{zero-bias} by using spline interpolation either on the biased Hamiltonian or set of current \textit{versus} bias points, if in number enough to do so. 
From these results, we extract the 2D conductivity, that is, the conductance normalized by the device area. 

\subsection{Pseudo-atomic orbital basis}
SIESTA is a DFT code that uses \mbox{pseudo-atomic} orbitals (PAOs) as a basis set. 
The benefit of using such a basis is twofold: (i) for large systems, the needed basis is smaller than the one needed for \mbox{plane-waves} (especially in the presence of large void spaces), and (ii) the localized basis set allows convenient transport calculations, as performed by TranSIESTA code. 
The downside is that it requires rigorous optimization to produce physical results. 
While increasing the \mbox{real-space} point density in a \mbox{plane-wave} code systematically improves the accuracy of the basis, that is not the case for a PAO basis. 
The basis optimization is performed, similarly to ref.~\cite{papior_simple_2018}, in order to reproduce the same band structure obtained with a \mbox{plane-wave} code in a $\pm$3~eV energy range around the Fermi level within an error of $\pm$100~meV. 
We use Abinit~{\cite{gonze_abinitproject_2020} as \mbox{plane-wave} code with, of course, the same pseudopotential and simulation parameters used for SIESTA code and given in the Appendix. 

\subsection{Strain}
The periodic unit cell required by DFT simulations entails strain in heterostructures with different lattice parameters. 
This is especially true in our case, because of the large lattice mismatch between the considered 2D materials and the metallic contacts. 
Such an artificial strain does not necessarily correspond to real experimental conditions. 
Therefore, we have defined a maximum allowed strain {\it per} material to have an average band deformation of $\pm$100~meV for Ni and $\pm$50~meV for 2D materials in a $\pm$2~eV range around Fermi level. 
As a consequence of this choice, the size of the supercell can considerably vary depending on the two materials. 
As shown in the next section, this entails a different sensitivity of the heterojunctions to different stacking modes.
The maximum strain for the materials considered is 1.4\%. 
A table of the final devices' supercell size and strain is given in the Appendix.

\section{Results and discussion}
This section shows our main results, which focus on different chosen 2D materials, their stacking modes with contacts, and number of layers. 
As introduced previously, we have selected three 2D materials, each with distinct electronic and structural properties, to examine their impact on the electronic properties at the interface with the electrodes and \mbox{out-of-plane} transport properties. 
For metallic electrodes, we consider Nickel, with the orientation (111), in Miller indices notation.

\subsection{The impact of stacking mode on the structural and electronic properties of Ni(111)$|$monolayer 2D material$|$Ni(111) heterostructures}

\begin{figure}[b]
     \centering
     \begin{subfigure}{1.0\textwidth}
         \centering
         \includegraphics[width=\textwidth]{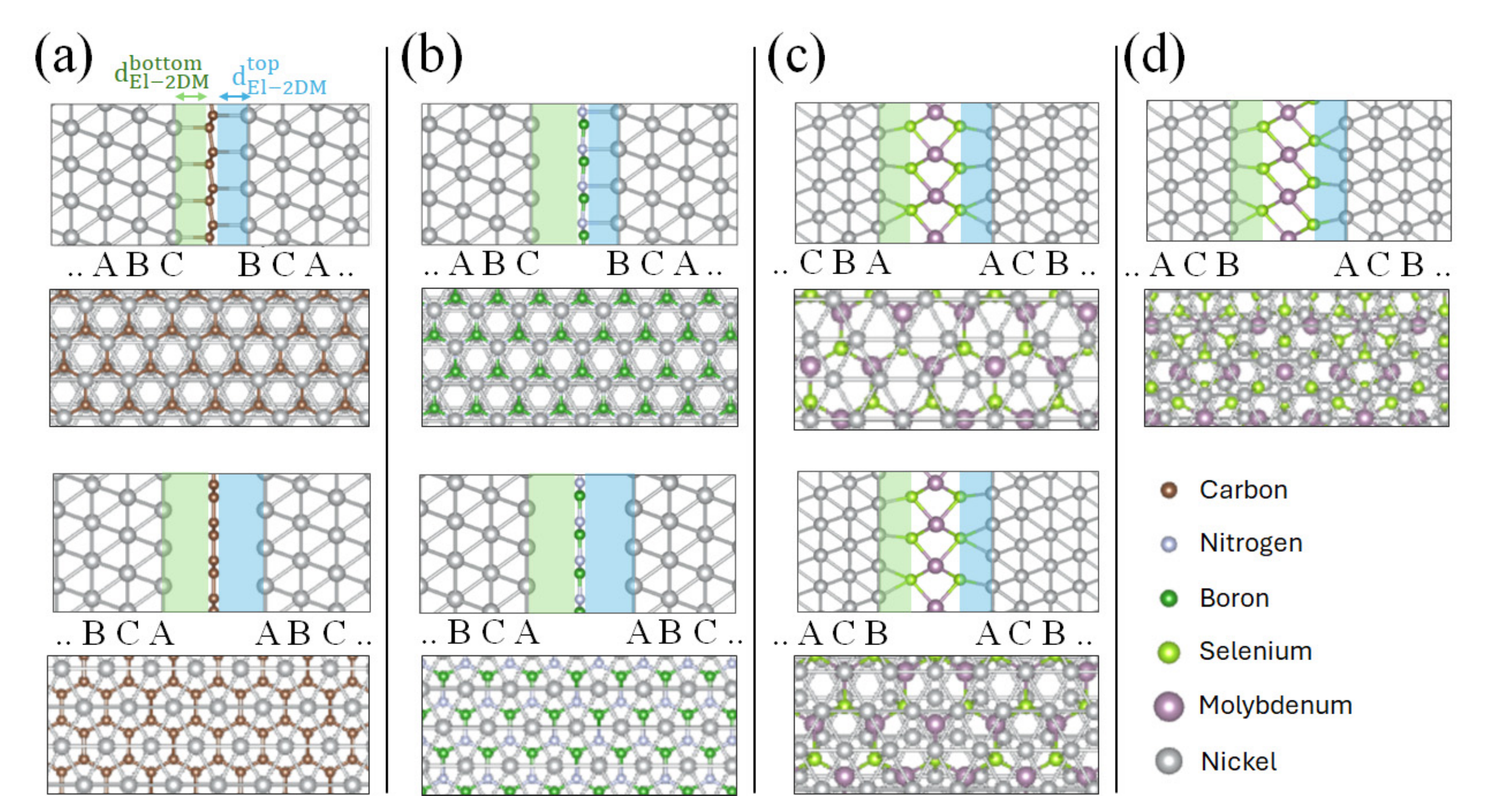}
     \end{subfigure}
        \caption{
          Lateral and top views of the structural configurations of (a)~\mbox{Ni$|$Gr$|$Ni}, (b)~\mbox{Ni$|$hBN$|$Ni}, (c)~\mbox{Ni$|$2H-MoSe$_2$$|$Ni} and (d)~\mbox{Ni$|$1T-MoSe$_2$$|$Ni}, for two different stacking modes. 
          The blue area represents the distance between the 2D material and the top electrode, while the green area represents the distance between the 2D material and the bottom electrode. 
          In the top views, only the 2D material and the first layer of each contact are shown.}
        \label{fig1}
\end{figure}

We start our study by investigating the impact of the stacking mode between Ni and the 2D materials on the electronic properties of their heterostructures. 

\Fref{fig1} shows the relaxed structures of the heterostructures \mbox{Ni$|$2D material$|$Ni}, where the 2D material can be Gr, hBN, \mbox{2H-MoSe$_2$} and \mbox{1T-MoSe$_2$}. 
The corresponding distances between the electrodes ($d_{\rm El-El}$), the top and bottom electrodes and the 2D material ($d_{\rm El-2DM}^{\rm top}$ and $d_{\rm El-2DM}^{\rm bottom}$, respectively), and the interaction energy are summarized in \tref{tab1}. 
Along the direction (111), Ni is composed of three layers (named A, B and C) with different stacking order and periodically repeated. 
The stacking modes of the heterojunctions are named according to the name of the Ni layers close to the 2D material, see \fref{fig1}. 
We have only considered two stacking modes among several possibilities for each 2D material, in order to discuss the significance of changing the layer stacking configuration. 
For graphene, see \fref{fig1}(a), the interfaces with the top and bottom electrodes are symmetric and exhibit an identical distance between the electrodes and the 2D material for the two stacking modes, i.e., $d_{\rm El-2DM}^{\rm top}=d_{\rm El-2DM}^{\rm bottom}$. 
In CB stacking mode, the system exhibits a higher interaction energy and shorter distances compared to those of stacking mode AA, because of the vertical alignment of C atoms with the closest Ni atoms. 
An analogous result is observed when the 2D material is hBN, see \fref{fig1}(b). 
In stacking mode AA, where B and N atoms are not aligned with Ni atoms, the $d_{\rm El-2DM}^{\rm top/bottom}$ distance is larger, the corrugation is less pronounced and the interaction energy is lower than in stacking mode BC. Here the interfaces are asymmetric, as hBN interacts more strongly with the top electrode where N atoms are vertically aligned with Ni atoms. 
In contrast, \mbox{Ni/2H-MoSe$_2$}, see \fref{fig1}(c), shows trivial changes depending on the stacking mode, as the Se atoms experience nearly identical Ni surroundings when averaged over the unit cell, thus leading to a similar interaction with top and bottom electrodes. 
This behavior, explained in detail later in this section, is related to the larger supercell size.
Hence, for the \mbox{Ni/1T-MoSe$_2$} interface, see \fref{fig1}(d), we considered only the BA stacking configuration. 

\begin{table}[t]
\caption{Interlayer distance and interaction energy for hBN, Gr, \mbox{1T-Mose$_2$} and \mbox{2H-MoSe$_2$ for two different stacking modes}.\label{tab1}}
\footnotesize\rm
\begin{tabular*}{\textwidth}{@{}l*{15}{@{\extracolsep{0pt plus12pt}}l}}
\br
Properties & Ni$|$Gr$|$Ni & Ni$|$hBN$|$Ni & Ni$|$2H-MoSe$_2|$Ni & Ni$|$1T-MoSe$_2|$Ni\\
\mr

   stacking mode  & AA | CB & AA | CB & AA | BA & BA\\
\mr

$d^{\rm bottom}_{\rm El-2DM}$ (\AA) & 2.85 | 2.08  & 3.02 | 3.02 & 2.26 | 2.25 & 2.19\\
$d^{\rm top}_{\rm El-2DM}$ (\AA) & 2.91 | 2.09  & 2.98 | 2.04 & 2.26 | 2.25 & 2.19\\
$d_{\rm El-El}$ (\AA) & 5.77 | 4.40 & 6.01 | 5.15 & 7.93 |  7.91 & 7.89\\
Interaction energy (meV/\AA$^2$) & 223 | 274 & 211 | 238 & 236 | 238 & 282\\
\br
\end{tabular*}
\end{table}

\begin{figure}[h!]
     \centering
     \begin{subfigure}{1.0\textwidth}
         \centering
         \includegraphics[width=\textwidth]{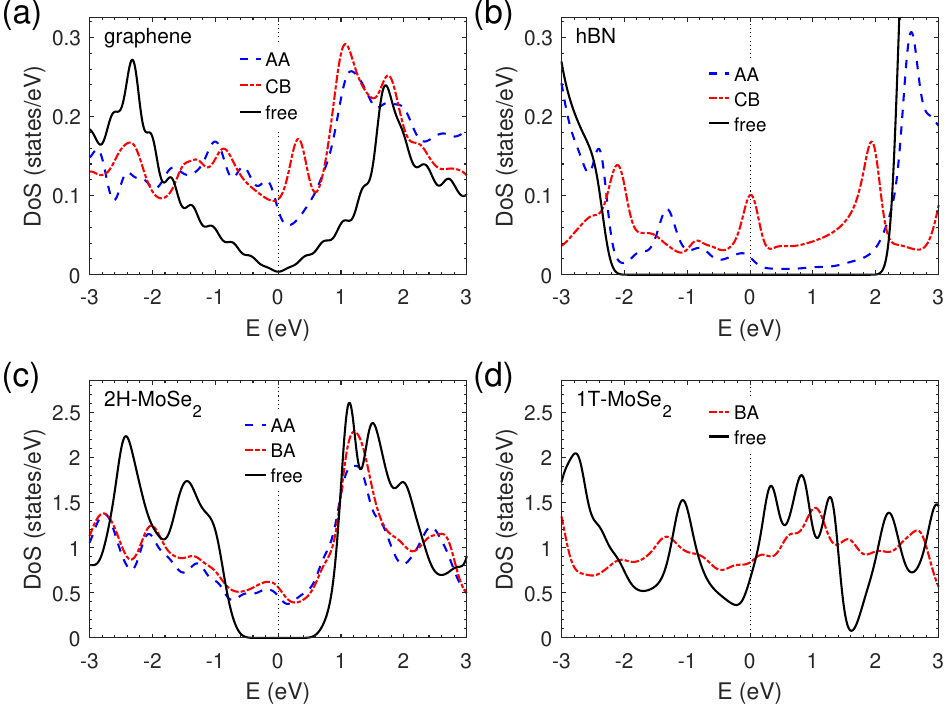}
     \end{subfigure}
        \caption{
          Total DoS {\it per} unit cell for monolayer (a) graphene, (b) hBN, (c) \mbox{2H-MoSe$_2$} and (d) \mbox{1T-MoSe$_2$} \textit{versus} energy (eV). 
          Fermi level at $E=0$ is indicated by a vertical dotted line. 
          Comparison between the 2D material relaxed in Ni electrodes with CB (or BA for \mbox{MoSe$_2$}) stacking (red dashed line), AA stacking (blue dashed line), and relaxed \mbox{free-standing} 2D material (black solid line).}
        \label{fig2}
\end{figure}

\begin{figure}[h!]
     \centering
         \centering
         \includegraphics[width=\textwidth]{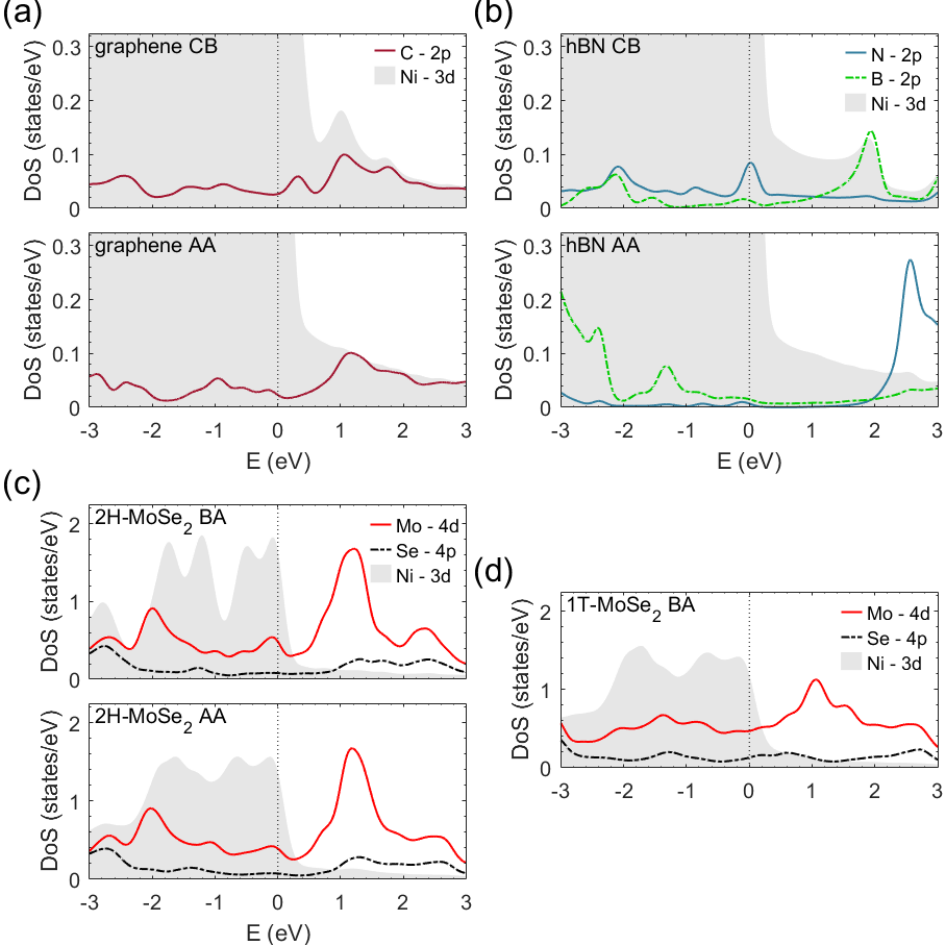}
        \caption{
          Projected DoS {\it per} atom on the indicated valence orbitals for monolayer (a) graphene, (b) hBN, (c) \mbox{2H-MoSe$_2$} and (d) \mbox{1T-MoSe$_2$} \textit{versus} energy and for different indicated stacking modes. 
          Fermi level at $E=0$ is indicated by a vertical dotted line. }
        \label{fig3}
\end{figure}

\begin{figure}[tb]
     \centering
     \begin{subfigure}{1.0\textwidth}
         \centering
         \includegraphics[width=\textwidth]{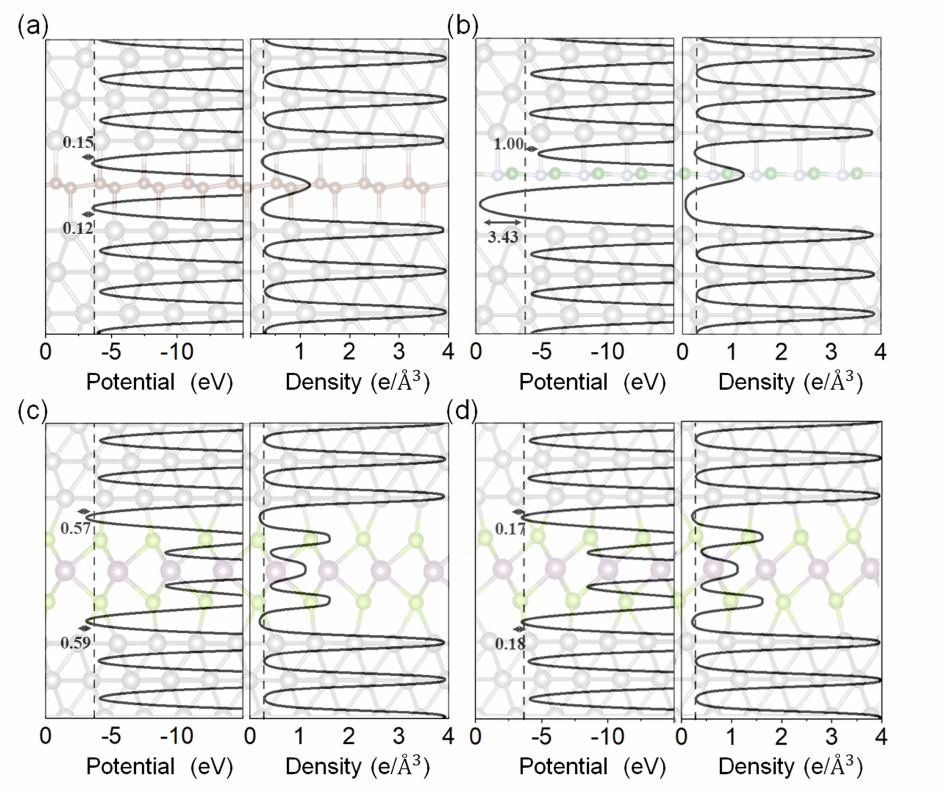}
     \end{subfigure}
        \caption{
         Planar average of the Hartree potential (left) and charge density (right) for monolayer (a) graphene, (b) hBN, (c) \mbox{2H-MoSe$_2$} and (d) \mbox{1T-MoSe$_2$} \textit{versus} \mbox{out-of-plane} distance (\AA). Dashed line represents Fermi level for the Hartree potentials and the minimum of charge density of electrodes. Black arrows represent the tunnelling barriers.}
        \label{fig4}
\end{figure}

The stacking arrangement impact on the atomic interactions at the contacts is observed in the electronic properties, as illustrated in \fref{fig2} by the total density of states (DoS), and in \fref{fig3} by the projected DoS. The DoS of the Ni atom plane closest to the 2D material is also provided, with coinciding peaks proving the hybridization of their orbitals. 
In the case of graphene, see \fref{fig2}(a), the interaction between the C and Ni atoms, which are aligned in the CB stacking mode, results in significant modifications of the electronic structure of graphene. 
Indeed, graphene becomes fully metallic as a result of the strongly hybridized states at the interface with the contacts. 
From \fref{fig2}(b), for the CB stacking mode, the DoS exhibits states within the bandgap of hBN. 
This is consequence of the orbital hybridization between the $p$ states of N atoms and the $d$ states of Ni atoms, as clearly visible in the top panel of \fref{fig3}(b), from the N DoS peak at the Fermi level. This finding is consistent with the previous observations that, based on interface distances, the \mbox{N-Ni} interaction is stronger than the \mbox{B-Ni} interaction, due to the higher electronegativity of N atoms compared to B atoms. 
On the other hand, in the stacking mode AA, the hBN/Ni interface exhibits a weaker interaction, which can be seen from the reduced impact on the hBN DoS see \fref{fig2}(b) and \fref{fig3}(b). 
These observations highlight the critical importance of atomic alignment in tuning the structural and electronic properties at interfaces, thus emphasizing the potential to engineer specific electronic characteristics through precise control of stacking configurations. 
For MoSe$_2$, see \fref{fig2}(c,d), both phases experience significant changes due to the interaction with the electrodes at the interface. 
More importantly, similarly to hBN, \mbox{2H-MoSe$_2$} loses its semiconducting behavior due to state hybridizations with Ni at the interfaces, thus leading to induced electronic states within the bandgap. 
As expected, \fref{fig2}(c) shows that the electronic properties exhibit less pronounced dependence on the stacking mode for \mbox{2H-MoSe$_2$}. 
The DoS remains relatively unchanged for the two stacking modes, which can be attributed to the use of a larger supercell composed of different numbers of unit cells of different materials, as \mbox{2H-MoSe$_2$(3$\times$3)/Ni(4$\times$4)}, see \fref{fig1}(c).
This results in a larger variety of relative atomic positions within the supercell and then in an arrangement in which the interfacial interactions are strong in some areas and weak in others, regardless of the stacking mode.} 
Consequently, the variations in stacking mode entail compensated interactions, which remain nearly similar for different stacking modes, when averaged over the supercell. 
This is in contrast with the cases of graphene and hBN, where smaller unit cells lead to direct contacts or weak interactions based on vertical alignment, see \fref{fig1}(a,b). 
Therefore, the trivial changes observed in the electronic properties of \mbox{2H-MoSe$_2$} with varying stacking modes arise from its differing interfacial environment, in addition to the fact that coupling between Se and Ni atoms is weaker than the N-Ni coupling discussed above. 
On the other hand, \mbox{1T-MoSe$_2$} keeps its metallic nature, see \fref{fig2}(d). 
In what follows, only energetically favorable stacking modes are considered for each heterostructure, which, according to the interaction energies reported in \tref{tab1}, are CB for graphene and hBN, and BA for MoSe$_2$.

To further elucidate the electronic behavior at the interface level, \fref{fig4} provides the \mbox{in-plane-averaged} charge distribution and the Hartree potential along the orthogonal axis, which explain the simulated conductivity reported in \tref{tab2}. 
\Fref{fig4}(a,c,d) shows that graphene and MoSe$_2$,  exhibit a symmetric charge distribution at the top and bottom interfaces. 
The effective potential shows a reduced tunnelling barrier at the interfaces with a higher charge accumulation and stronger interaction. 
In particular, graphene exhibits the lowest tunnelling barrier, which facilitates electron transfer across the interface and increases the transmission probability, thereby enhancing the overall conductivity of the system. 
Accordingly, the highest electrical conductivity is observed in the Ni$|$Gr$|$Ni structure, due to the minimal tunnelling barrier.
For the hBN/Ni interface, see \fref{fig4}(b), the interaction with the top electrode, where \mbox{N-Ni} bonding occurs, shows a charge density higher than at the bottom interface. 
This arises from the interaction of N atoms with Ni atoms, which is stronger than that of B atoms with Ni atoms, thus resulting in an asymmetric charge distribution.  
The lowest electrical conductivity of the Ni$|$hBN$|$Ni heterostructure results from the highest tunnelling barrier at the interface dominated by the weaker B-Ni interaction. 

Indeed, based on our results, the vertical transmission and the trend of the conductivity is related to the heights of the tunnel barriers, 2D material DoS at Fermi level, the distance between the electrodes and the 2D material, as well as its thickness.  
The high conductivity in graphene is consistent with its low tunnelling barrier, as mentioned above, and increased DoS, which enhances the electron transmission across the interface. 
In addition, the short $d_{\rm El-2DM}^{\rm top/bottom}$ supports stronger interface coupling and facilitates charge transfer. 
Compared to graphene and despite having close tunnelling barrier heights, the lower conductivity through \mbox{1T-MoSe$_2$} may be attributed to its larger thickness. 
The \mbox{2H-MoSe$_2$} configuration shows slightly lower conductivity ($< 0.1$~$\mu$S/${\rm \AA}^2$) than \mbox{1T-MoSe$_2$}, due to its approximately 0.4~eV higher tunnelling barriers while having almost the same thickness and relatively close state densities at the Fermi level.
Finally, despite being thinner than MoSe$_2$, hBN shows the lowest conductivity, due to the significantly higher tunnelling barrier of 3.43~eV. 
The explanation can be found in the asymmetric tunnelling barriers, with one barrier being very high, thus suggesting that electron transmission is significantly blocked at one interface, thus resulting in reduced overall conductivity. 

We observe significant modifications in the electronic characteristics of all 2D materials studied, due to their interaction with \mbox{non-passivated} electrodes. 
In particular, they become metallic when they previously were semiconducting or insulating. 
As a consequence, the transmission and electrical conductivity of their heterostructures are not intrinsically based on the electronic properties of the isolated materials. 
In the next section, we consider multilayer 2D materials to observe the impact of the intrinsic electronic properties on the electrical conductivity. 

\begin{table} [tb]
\caption{ \label{tab2}
   Conductivity of hBN, Gr, \mbox{1T-MoSe$_2$} and \mbox{2H-MoSe$_2$} monolayers with Ni contacts. 
   The conductivity is normalized by the surface of the device. 
   }
\footnotesize\rm
\begin{tabular*}{\textwidth}{@{}l*{15}{@{\extracolsep{0pt plus12pt}}l}}
\br
 Heterostructures & Ni$|$hBN$|$Ni & Ni$|$2H-MoSe$_2|$Ni & Ni$|$1T-MoSe$_2|$Ni & Ni$|$Gr$|$Ni\\
\mr
conductivity ($\mu$S/\AA$^2$) & 1.33 & 1.88 & 1.95 & 3.27 \\
\br
\end{tabular*}
\end{table}

\subsection{Electronic and transport properties of Ni(111)$|$multilayer 2D material$|$Ni(111) heterostructures}

\begin{figure}[b]
     \centering
      \includegraphics[width=\textwidth]{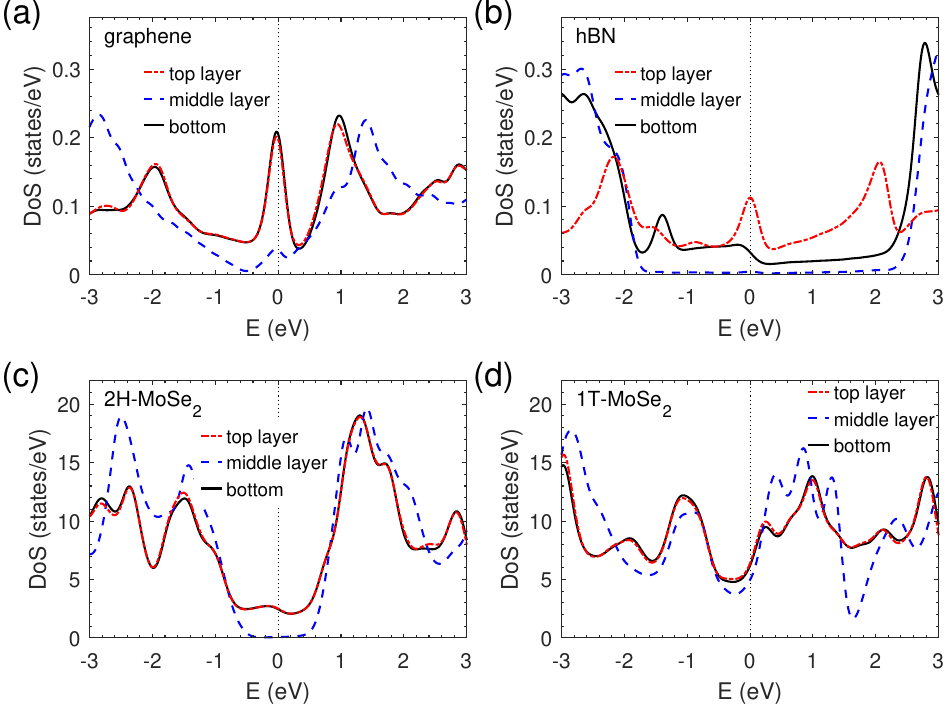}
        \caption{
          DoS on the layers in contact with the top/bottom electrodes and the middle layer for trilayer (a)~graphene, (b)~hBN, (c)~\mbox{2H-MoSe$_2$} and (d)~\mbox{1T-MoSe$_2$} \textit{versus} energy (eV). 
          Fermi level at $E=0$ is indicated by a vertical dotted line.}
        \label{fig5}
\end{figure}

\begin{table}[b]
\caption{ \label{tab3}
  Conductivity for multilayer hBN, Gr, \mbox{1T-MoSe$_2$} and \mbox{2H-MoSe$_2$} for different number of layers.}  
\begin{tabular*}{\textwidth}{@{}l*{15}{@{\extracolsep{0pt plus12pt}}l}}
\br
conductivity & Ni$|$Gr$|$Ni & Ni$|$hBN$|$Ni & Ni$|$2H-MoSe$_2|$Ni & Ni$|$1T-MoSe$_2|$Ni \\
\mr
ML ($\mu$S/\AA$^2$) & 3.27 & 1.33 & 1.88 & 1.95 \\
2L ($\mu$S/\AA$^2$)& 0.48 & 0.20 & 0.06 & --\\
3L ($\mu$S/\AA$^2$)& 7.0$\times$10$^{-3}$ & 2.0$\times$10$^{-3}$ & 2.0$\times$10$^{-3}$ & 0.24 \\
4L ($\mu$S/\AA$^2$)& 1.4$\times$10$^{-4}$ & 9.4$\times$10$^{-5}$ & 5.8$\times$10$^{-5}$ & -- \\
\br
\end{tabular*} 
\end{table}

\begin{figure}[h!]
     \centering
     \begin{subfigure}{1.0\textwidth}
         \centering
         \includegraphics[width=\textwidth]{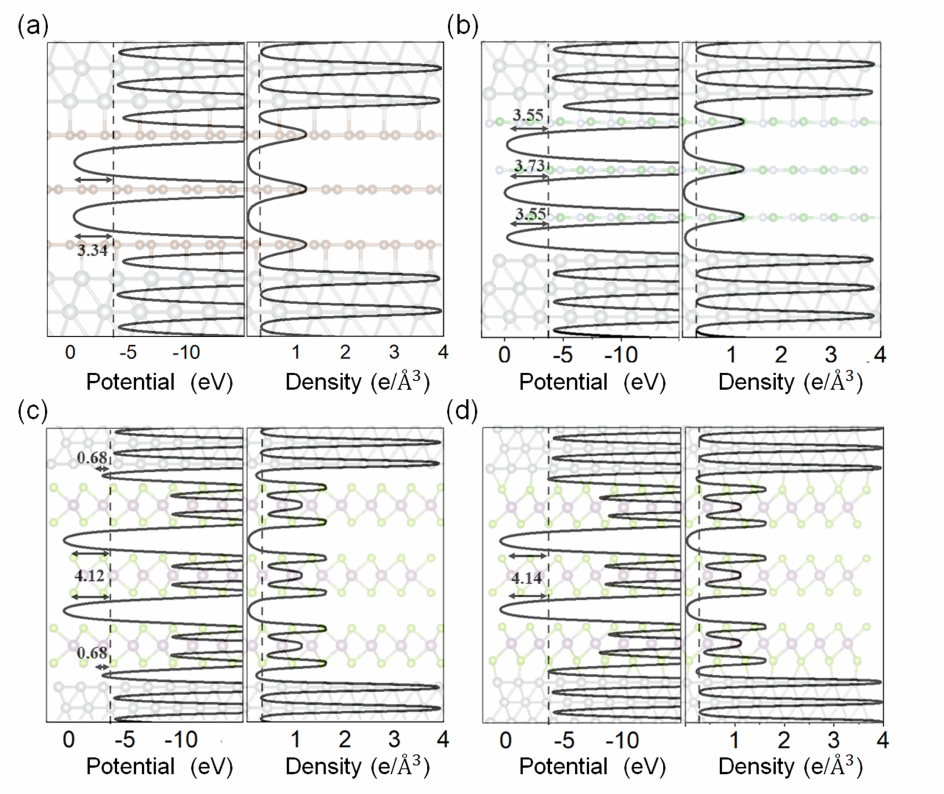}
     \end{subfigure}
        \caption{
         Planar average of the Hartree potential (left) and charge density (right) for trilayer (a) graphene, (b) hBN, (c) \mbox{2H-MoSe$_2$} and (d) \mbox{1T-MoSe$_2$} \textit{versus} \mbox{out-of-plane} distance (\AA). Dashed line represents Fermi level for the Hartree potentials and the minimum of charge density of electrodes. Black arrows represent the tunnelling barriers.}
        \label{fig6}
\end{figure}

In this section, we consider multilayer 2D materials, among which, trilayers represent the thinnest multilayer structure that distinguishes the behavior between the middle layer and the layers close to the metallic contacts.  
This approach allows us to examine the individual contributions of the interfacial layers and the \mbox{non-interacting} middle layer, to the electronic and transport properties of the system as a whole. 
The corresponding DoS of our heterostructures with different numbers of \mbox{2D-material} layers is reported in \fref{fig5}. The figure shows the DoS projection on the different layers of a trilayer, which reveals significant differences in their electronic structure. 
For all four 2D materials investigated, the layer in direct contact with the electrodes exhibits electronic characteristics similar to those in the monolayer cases, due to strong orbital hybridization with the metal atoms at the interface. 
For graphene, \mbox{2H-MoSe$_2$} and \mbox{1T-MoSe$_2$}, due to the symmetric top and bottom \mbox{layers-electrode} distances, i.e., $d_{\rm El-2DM}^{\rm top}=d_{\rm El-2DM}^{\rm bottom}$, the DoS is the same on the top and bottom layers, except for small numerical errors, see \fref{fig5}(a,c,d). 
For hBN, the different \mbox{layers-electrode} distances, with $d_{\rm El-2DM}^{\rm top}<d_{\rm El-2DM}^{\rm bottom}$, entail a difference between the DoS in the top and bottom layers. 
In fact, as shown in \fref{fig5}(b), the stronger bond between the top electrode and the hBN layer, evidenced by the peak at the Fermi energy, leads to a greater DoS than that on the bottom layer.
In contrast, the middle layer, see \fref{fig5}, is significantly less influenced by the metallic contacts and preserves its intrinsic electronic structure. 
This limited interaction allows the middle layer to maintain characteristics closer to its isolated form, apart from an energy shift induced by doping due to the charge transfer. 
In fact, as presented in \fref{fig6}, the reduced potential barrier and increased charge density are located only at the 2D material/Ni interface. 
In contrast, the interfaces between the 2D material layers show no significant charge density accumulation and exhibit high tunnelling barriers, which highlights the role of the middle layer in reducing the tunnelling current through the heterostructure. 
Then, the 2D contact layers act as passivating layers for the metallic contacts. 
As a result, the conductivity, see \tref{tab3}, decreases when increasing the number of layers. 
This trend is consistently observed over all the studied 2D materials. 
The trilayer \mbox{1T-MoSe$_2$} shows a significantly higher conductivity value compared to the other 2D materials, which is due to its higher DoS at the Fermi level, especially in the middle layer, see \fref{fig5}(d). 
This contrasts with the semiconducting behavior of its 2H counterpart and hBN. 
These findings highlight the critical role played by the intrinsic electronic properties of 2D material in the overall transport behavior in multilayered 2D material systems.
In the case of semiconducting 2D materials, we expect the conductivity to go to zero when increasing the thickness. 
In the case of metallic 2D materials, we expect the conductivity to saturate to a finite limit, mostly determined by the interlayer coupling and lower than its in-plane counterpart. 
In any case, according to our results, the conductivity is not strictly zero even with four layers and, with three layers, it is still much higher than experimentally observed~\cite{Ge2017}. 
This suggests the presence of important experimental characteristics not considered here, such as multilayer regions, polymer residues, or metal oxidation, which could bring the system deeper into the tunnelling regime and suppress its conductivity by orders of magnitude.

\section{Conclusion}
This work demonstrates the influence of stacking configurations and material thickness on the electronic and transport properties of metal$|$2D material$|$metal systems. 
In monolayer configurations, strong interactions with metal electrodes result in largely altered electronic states with the formation of gap states. 
However, in trilayer systems, the middle layer maintains its intrinsic properties due to the limited electrode interaction. 
Among the systems studied, graphene exhibits the highest conductivity due to its low tunnelling barrier and increased DoS at the Fermi level, while hBN exhibits minimal conductivity due to its high tunnelling barrier. We conclude that the \mbox{out-of-plane} transport characteristics are not directly related to the \mbox{in-plane} ones. 
In particular, a single hBN monolayer within \mbox{non-passivated} metallic contacts turns out to be conductive, despite its insulating nature when isolated. 
Our findings are of particular relevance for the use of 2D materials in the field of beyond CMOS electronics, with a specific emphasis on the development of atomristors.

\section*{Acknowledgements}
This work is supported by the ANR SWIT project (ANR-19-CE24-0004) and the LabEx Minos (ANR-10-LABX-55-01). 
It is implemented using HPC resources from GENCI–IDRIS (Grant 2023 A0140914157), GENCI-TGCC (Grant 2024 A0160914157) and the GRICAD infrastructure (https://gricad.univ-grenoble-alpes.fr), which is supported by Grenoble research communities.

\newpage
\appendix
\section*{Appendix}
\setcounter{section}{1}
\begin{table} [h!]
\caption{
   SIESTA basic simulation parameters: real-space density grid cutoff (\mbox{R-grid}) and the reciprocal space grid (\mbox{K-grid}) Monkhorst-Pack matrix. 
   The reciprocal space grid is increased for transport calculations (TRANSIESTA and TBtrans).
   }
\footnotesize\rm
\begin{tabular*}{\textwidth}{@{}l*{15}{@{\extracolsep{0pt plus12pt}}l}}
\br
Devices & Ni$|$hBN$|$Ni & Ni$|$2H-MoSe$_2|$Ni & Ni$|$1T-MoSe$_2|$Ni & Ni$|$Gr$|$Ni\\
\mr
SIESTA K-grid & 20$\times$20$\times$3 & 5$\times$5$\times$1 & 6$\times$6$\times$1 & 20$\times$20$\times$3 \\
TRANSIESTA K-grid & 40$\times$40$\times$1 & 10$\times$10$\times$1 & 12$\times$12$\times$1 & 40$\times$40$\times$1  \\
TBtrans K-grid & 80$\times$80$\times$1 & 25$\times$25$\times$1 & 24$\times$24$\times$1 & 80$\times$80$\times$1  \\
R-grid (Ry)  & 1000 & 1000  & 1000  & 1000 \\
\br
\end{tabular*}
\end{table}
\begin{table} [h!]
\caption{
      Basis parameters. 
      The basis size is given in the conventional formalism with a \mbox{double-zeta} basis (DZ) with either a polarized orbital (P), two (DP) or a single \mbox{non-polarized} orbital (nP). 
      The energy shift sets all the radii of the first radial functions. 
      The smaller the energy shift, the longer the ranges of the first radial functions. 
      The splitnorm factor is defined for each second radial function. 
      The smaller the splitnorm, the closer the second radial function is to the first one. 
      $R_{c1}$ and $R_{c2}$ are the \mbox{cut-off} radii of the first and second radial functions for specific orbitals. 
      The number of orbitals in the unit cell is given by the row \textit{orbitals}.
      }
\footnotesize\rm
\begin{tabular*}{\textwidth}{@{}l*{15}{@{\extracolsep{0pt plus12pt}}l}}
\br
Material & Ni & hBN & 2H-MoSe$_2$ & 1T-MoSe$_2$ & Gr\\
\mr
Basis size & DZDP & DZP+3s+3p nP & DZ + 4f P & DZP + 4f P & DZP+3s+3p nP \\
Energy shift (meV) & 110 & 110 & 70 & 70 & 90 \\
Splitnorm & 4s:0.6 & N2s:0.5  & 5s:0.6  & 5s:0.4 & 2s:0.25\\
& 3d:0.15 &  N2p:0.5  & 4d:0.1 & 4d:0.25  &  2p:0.2\\
& & B2s:0.5 & 3s:0.05 & 3s:0.35 & \\
& & B2p:0.5 & 3p:0.3 & 3p:0.35 & \\
$R_{c1}$ (Bohr) & & B3s,p:9.5 & & & 3s:9  \\
$R_{c2}$ (Bohr) & & N3s,p:8.0 & & & 3p:9 \\
Orbitals & 22 & 34 & 59 & 62 & 34 \\
\br
\end{tabular*}
\end{table}
\begin{table} [h!]
\caption{
    Supercell size and found strains for each device. 
    The supercell size is given by the number of unit cells of the 2D material times the number of unit cells of the electrode metal. 
    Example: a 3$\times$4 device supercell means a 3$\times$3 2D supercell and a 4$\times$4 electrode supercell.
    }
\footnotesize\rm
\begin{tabular*}{\textwidth}{@{}l*{15}{@{\extracolsep{0pt plus12pt}}l}}
\br
Properties & Ni$|$hBN$|$Ni & Ni$|$\mbox{2H-MoSe$_2$}$|$Ni & Ni$|$\mbox{1T-MoSe$_2$}$|$Ni & Ni$|$Gr$|$Ni\\
\mr
Supercell & 1$\times$1 & 3$\times$4 & 3$\times$4 & 1$\times$1 \\
Strain 2D (\%) & 0.56  & 0.03 & 0.0 & 0.06\\
Strain Ni (\%) & 1.40  & 0.44  & 0.16 & 0.27\\
\br
\end{tabular*}
\end{table}

\newpage
\section*{References}

\providecommand{\newblock}{}

\end{document}